\documentclass[aps, prl, twocolumn, showpacs,amsmath,amssymb,superscriptaddress]{revtex4}
\usepackage{latexsym}
\usepackage{amssymb}
\usepackage{epsfig,amsmath,graphics}
\usepackage{epstopdf}

\begin{document}


\title{A New Method for Gravitational Wave Detection with Atomic Sensors}

\author{Peter W. Graham}
\affiliation{Stanford Institute for Theoretical Physics, Department of Physics, Stanford University, Stanford, CA 94305}

\author{Jason M. Hogan}
\affiliation{Department of Physics, Stanford University, Stanford, CA 94305}

\author{Mark A. Kasevich}
\affiliation{Department of Physics, Stanford University, Stanford, CA 94305}

\author{Surjeet Rajendran}
\affiliation{Stanford Institute for Theoretical Physics, Department of Physics, Stanford University, Stanford, CA 94305}

\date{\today}

\begin{abstract}
Laser frequency noise is a dominant noise background for the detection of gravitational waves using long-baseline optical interferometry.  Amelioration of this noise requires near simultaneous strain measurements on more than one interferometer baseline, necessitating, for example, more than two satellites for a space-based detector, or two interferometer arms for a ground-based detector.  We describe a new detection strategy based on recent advances in optical atomic clocks and atom interferometry which can operate at long-baselines and which is immune to laser frequency noise.  Laser frequency noise is suppressed because the signal arises strictly from the light propagation time between two ensembles of atoms.  This new class of sensor allows sensitive gravitational wave detection with only a single baseline. This approach also has practical applications in, for example, the development of ultra-sensitive gravimeters and gravity gradiometers.
\end{abstract}

\pacs{04.80.-y, 04.80.Nn, 95.55.Ym, 03.75.Dg}
\maketitle


The observation of gravitational waves will open a new spectrum in which to view the universe \cite{Schultz}.  Existing detection strategies are based on long-baseline optical interferometry \cite{Abbott:2007kv, Accadia:2012zz}, where gravitational waves induce time-varying phase shifts in the optical paths.  Spurious phase shifts arising from laser frequency and phase noise are suppressed through multi-arm configurations which exploit the quadrupolar nature of gravitational radiation to separate gravitational wave induced phase shifts from those arising from laser noise.  In the absence of such noise, a single baseline optical interferometer, {\it e.g.}~a Fabry-Perot interferometer, would suffice for gravitational wave detection.  In these detectors, stringent constraints are also placed on the mechanical motion of the interferometer optics in order to avoid optical path length fluctuations which would otherwise obscure the gravitational wave signals.

We propose a new approach, based on recent advances in optical frequency control and atom interferometry, which directly avoids laser frequency noise and naturally mitigates mechanical noise sources.  The approach draws on the development of light-pulse gravity gradiometers, where Doppler-sensitive two-photon optical transitions are used to measure the differential acceleration of two spatially separated, free-falling, laser cooled atomic ensembles \cite{Snadden, McGuirk, Fixler}.  For these sensors, the optical interrogation is configured so that the same laser beams interrogate both ensembles of atoms along a common line-of-sight.  This significantly suppresses laser frequency noise, but does not remove it completely due to the time delay introduced by the travel time of the light between ensembles and the need for each of the two counter-propagating laser beams to temporally overlap (in order to drive the two-photon transitions) \cite{McGuirk, Biedermann}.  For shorter baseline instruments ({\it e.g.} 1 m gravity gradiometers), this noise source is relatively benign.  For longer-baseline gravitational wave detectors ({\it e.g.} 10 km - 1000 km baseline AGIS proposals described in Refs. \cite{Dimopoulos:2008sv, Hogan:2010fz}), it becomes a dominant noise source  \cite{Baker}.  It also places stringent limits on knowledge of residual accelerations of the laser platform, which manifest themselves as Doppler shifts on the frequency of the light in the inertial frame of the atoms.

Laser noise would nearly disappear if the atomic transitions were driven with a single laser pulse since the laser frequency noise in each pulse would be common to both atom interferometers and would cancel in the differential measurement.  This follows from the relativistic formulation of atom interferometry in Ref.~\cite{Dimopoulos:2008hx, Dimopoulos:2006nk} since the laser phase of a pulse is set when the pulse is emitted and does not change as it propagates along the null geodesic connecting the laser to the atoms.  We propose a laser excitation protocol which is based solely on single photon transitions in order to exploit this noise immunity and which is capable of achieving scientifically interesting strain sensitivities.  In an optical interferometric gravitational wave detector, the relative phases of the interfering optical fields serve as proxies for the propagation time of the light along the interferometer arms.  In the proposed approach,  gravitational waves are instead sensed by direct measurement of the time intervals between optical pulses, as registered by atomic transitions which serve as high stability oscillators.

\begin{figure}
\begin{center}
\includegraphics[width=\columnwidth]{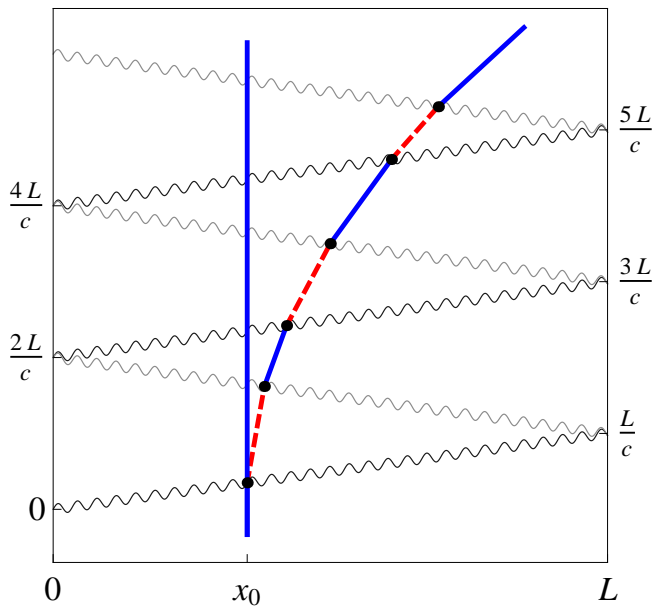}
\caption{ \label{Fig:beamsplitter} A space-time diagram of our proposed LMT beamsplitter with $N=3$. The solid (blue) lines indicate the motion of an atom in the ground state, the dashed (red) lines indicate the atom in the excited state. Light pulses from the primary and secondary lasers are incident from the left (dark gray) and the right (light gray) respectively.  Dots indicate the vertices at which the laser interacts with the atom.}
\end{center}
\end{figure}

{\it A New Type of Atom Interferometer}--  Due to atomic momentum recoil in the absorption and stimulated emission of photons during optical interactions, the proposed pulse sequence, detailed below, can be understood as a variant of a light-pulse de Broglie wave interferometer in a Mach-Zender configuration \cite{Borde, Helmcke, KasevichChu}.  A prototypical excitation sequence can be described as a combination of beamsplitter and mirror segments.

For the beamsplitter, the lasers are pulsed as in Fig.~\ref{Fig:beamsplitter}.  The primary laser is taken to be at $x=0$, the left side of the figure, the secondary laser is taken at $x=L$, the right side of the figure.  The atom begins at $x=x_0$ in the ground state.  The initial pulse at time $t=0$ is a $\pi / 2$ pulse which splits the atom's wavefunction in two (for simplicity, we neglect spontaneous emission from the excited state).  Some time after this reaches $x=L$, a $\pi$ pulse is fired from the secondary laser which is Doppler tuned to interact only with the half of the atomic wavefunction which was originally excited.  In Fig.~\ref{Fig:beamsplitter} the second pulse is taken to leave at the time $L / c$ when the first pulse arrives at $x=L$, but in fact it is only necessary that the second pulse leaves after this time.  After the initial pair of pulses, to make a large momentum transfer (LMT) beamsplitter $N-1$ more pairs of $\pi$ pulses are sent, each pair having the first pulse from the primary laser and the second from the secondary laser.  The frequency of these pulses are tuned so they interact only with the faster half of the atom.  This is shown in Fig.~\ref{Fig:beamsplitter} for $N=3$.  This leaves half of the atom's wavefunction in the ground state with unchanged momentum (the left solid line in Fig.~\ref{Fig:beamsplitter}) and gives a momentum of $2 N \hbar k$ to the other half of the atom, where $k$ is the wavevector of each pulse.  This sequence makes an LMT beamsplitter using only single-photon atomic transitions.  Note that according to the standard rules which govern the laser/atom interactions, the phase of the laser field is read into the atomic coherence during each of the atomic transitions.

\begin{figure}
\begin{center}
\includegraphics[width=\columnwidth]{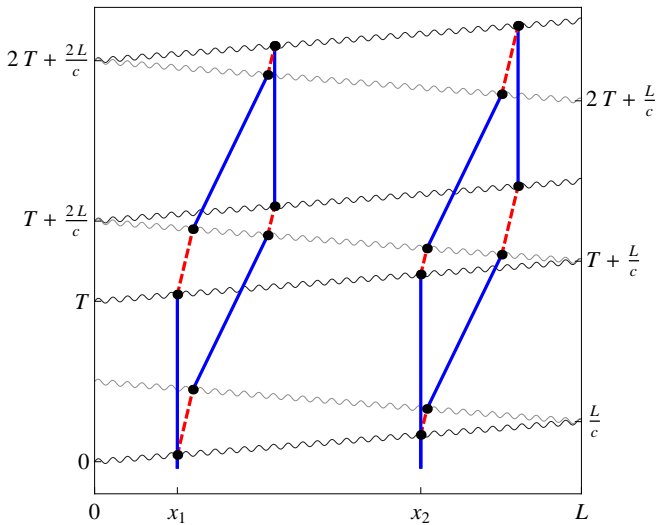}
\caption{ \label{Fig:setup} A space-time diagram of the proposed configuration of a differential measurement between two atom interferometers beginning at positions $x_1$ and $x_2$. The lines are as in Fig.~\ref{Fig:beamsplitter}.  For clarity the beamsplitters shown are not LMT, {\it i.e.}~here $N=1$.}
\end{center}
\end{figure}

The basic mirror sequence is three $\pi$ pulses, alternately from the primary and secondary lasers as shown in the middle of Fig.~\ref{Fig:setup}.  In general, there are several ways to realize this sequence.  It can begin either from the primary laser (as shown in Fig.~\ref{Fig:setup}) or from the secondary laser.  The pulses are tuned to interact only with certain halves of the atom, as indicated by the dots in Fig.~\ref{Fig:setup}.  To make the entire LMT mirror pulse, $N-1$ pairs of laser pulses are added before the basic mirror sequence to slow down the fast half of the atom, the exact opposite of the initial beamsplitter.  Similarly $N-1$ pairs are added after the basic mirror sequence to accelerate the other half of the atom.  This reverses the momenta of the two incoming halves of the atom's wavefunction.  The slow half gets a momentum kick of $2 N \hbar k$, the fast half loses $2 N \hbar k$.

Using a beamsplitter-mirror-beamsplitter sequence allows the atom interferometer to close, so that the two halves of the atom's wavefunction overlap at and can be interfered by the final beamsplitter.  The phase difference is read out by measuring the atom populations in the interferometer output ports.  The mirror pulse is started at time $t=T$ and the final beamsplitter is started at time $t = 2T + \frac{L}{c}$.  This is shown in each half of Fig.~\ref{Fig:setup}.

This type of atom interferometer acts effectively as an accelerometer.  If the atom does not accelerate, the time spent in the excited state is the same for each half of the atom's wavefunction and there is no phase difference.  However if the atom accelerates, this time is not the same.  Since the atom accumulates phase faster in the excited state, this gives rise to a phase shift proportionally to the acceleration.  Interestingly, the phase shift is read in to the atom during the relatively short beamsplitter and mirror sequences themselves, not during the large interrogation time $\sim T$ between them.  Nevertheless, these phase shifts scale proportionally to $T$ since they depend on the change in the light travel time across the baseline between the beamsplitter and mirror sequences.  The phase shift (or sensitivity) of this type of atom interferometer also scales with $N$.  The leading order phase shift in a local gravitational field is $\sim N \omega_a g T^2/c$ where $\omega_a$ is the atomic energy level difference and $g$ is the acceleration due to gravity (here assumed constant in space and time).  The phase shift due to a gravitational wave is approximately the same with $g$ replaced by the acceleration caused by the gravitational wave.  Intuitively the factor of $N$ arises because the signal comes from the extra time spent in the excited state [the dashed (red) lines in Fig.~\ref{Fig:setup})] which increases linearly with $N$.

These leading order phase shifts are proportional to the atomic energy difference $\omega_a$, not to the laser frequency $\omega = kc$.  This is a known difference between atom optics based on two-photon Raman or Bragg transition (where $\omega_a \ll 1$ eV), and a single-photon transition (where $\omega_a$ is large, $\sim 1$ eV) \cite{Dimopoulos:2008hx}.  In practice the laser must be tuned so that $\omega$ is close to $\omega_a$ in order to drive the atomic transition.

{\it A Differential Measurement}--  A single interferometer of the type described above will have laser noise, but this can be removed by a differential measurement between two such interferometers (similar to the scheme proposed in Refs. \cite{Dimopoulos:2007cj, Dimopoulos:2008sv,Hogan:2010fz}).  The primary and secondary lasers are separated by a large distance $L$, with atom interferometers operated near them.  The atom clouds are initially prepared as described in \cite{Dimopoulos:2008sv}.  These two widely separated atom interferometers are run using common laser beams (see Fig.~\ref{Fig:setup}) and their differential phase shifts measured.  Importantly, for any given interogation, the same laser beam drives both interferometers. For example, the pulse from the primary laser at time $t=0$ triggers the initial beamsplitter for both interferometers and the pulse from the secondary laser at time $t=L/c$ completes this beamsplitter, again for {\it both} interferometers.  We will show that the differential phase shift between these interferometers contains a gravitational wave signal proportional to the distance between them. However, since the {\it same} laser pulse operates {\it both} interferometers, the differential signal is largely immune to laser frequency noise.  This idea has some similar features to the proposal described in Ref. \cite{Yu}, where a single laser only is used to interrogate two spatially separated atomic ensembles.

To see the effect of a gravitational wave on the differential phase between the two interferometers, assume that one interferometer is at $x_1 = 0$ in Fig.~\ref{Fig:setup} while the other is at $x_2 = L$ and $T \gg L/c$. In the absence of a gravitational wave, each arm spends a time $L/c$ in the excited state leading to a null result in each interferometer. Note though that the arms of the interferometer at $x_1$ spend time $L/c$ in the excited state in the beginning and the middle of the interferometer, while the arms of the interferometer at $x_2$ spend time $L/c$ in the excited state in the middle and end (see dashed lines in Figure \ref{Fig:setup}). In the presence of a gravitational wave of strain $h$ and frequency $\omega$, the distance between the atom interferometers oscillates in time.  This affects the laser pulse travel time which in turn affects the relative time spent by each atom interferometer arm in the excited state (see Fig.~\ref{Fig:setup}).
When $T \sim 1/\omega$ the distance changes by $\sim h L$ in time $T$ (assuming $\omega L/c \ll 1$).
Hence, the two interferometers spend a slightly different amount of time $\sim h \frac{L}{c}$ in the excited state.
This leads to a differential phase shift between the interferometers of $\sim \omega_a h L/c$.
For an LMT sequence with $N$ pulses, the phase shift is enhanced by $N$ since it adds during each pulse.  A fully relativistic calculation following the formalism of \cite{Dimopoulos:2008hx} yields the differential phase shift to be
\begin{equation}
\Delta \phi = \frac{4  N  \omega_a h}{c} \left(x_1 - x_2\right) \sin^{2} \left( \frac{\omega \, T}{2}\right) \sin \left(\phi_0 + \omega \, T \right)
\label{Eq:PhaseShift}
\end{equation}
proportional to the baseline $x_1 - x_2 \approx L$. $\phi_0$ in this expression is  the phase of the gravitational wave at the start of the experiment, whose change ($\phi_0 = \omega  t_0$) causes a time dependent phase shift in the experiment.

The gravitational wave signal is due to the oscillation of the laser ranging distance between the two interferometers.  The atoms effectively measure the light travel time across the baseline.  Thus, the lasers do not serve as a clock and so do not need a highly stable phase evolution.  Remarkably, only the constancy of the speed of light across the baseline is relevant.  This is an important change from all other interferometric gravitational wave detection schemes, where the laser serves the role of a phase reference, thus requiring additional noise mitigation strategies ({\it e.g.}~additional measurement baselines).

{\it Backgrounds}--  We will now discuss possible noise sources for the proposed scheme. We distinguish between two classes of noise: intrinsic laser noise and kinematic noise. Intrinsic laser noise refers to jitters in the phase and frequency of the laser while  kinematic noise is caused by the acceleration noise of the laser platform and jitter in the timing between the interferometer pulses. The phase of a laser pulse does not evolve during its propagation in vacuum from the laser to the location of the atom \footnote{Noise can arise from fluctuations in the refractive index $n$ of the medium of light propagation.  For space-based detectors, fluctuations $\delta n_p$ in the solar plasma density $n_p$ give an effective strain of $h \sim (n - 1) \delta n_p / n_p \lesssim \frac{10^{-22}}{\sqrt{\text{Hz}}}$, which would not limit the proposed detector \cite{Neugebauer}.}. Hence the atoms record the phase of the laser which exists at the emission time of the pulse.  Since both interferometers are operated by the same laser pulses, the intrinsic laser noise read by both interferometers is identical and will cancel in the differential phase.  The kinematic sources of noise affect both the imprinted laser phase and the amount of time spent by the arms of the interferometer in the excited state. Again, the noise from the imprinted laser phase will completely cancel in the differential measurement since the same laser pulses are used to drive both interferometers.  However, any kinematic difference such as a relative velocity $\Delta v$ between the two interferometers will result in differences in the time spent in the excited state between the two interferometers, leading to a differential phase shift suppressed by $\frac{\Delta v}{c}$.

\begin{table}
\renewcommand{\arraystretch}{1.5}
\begin{center}
\begin{tabular}{|l|c|c|l|}
\hline
& Phase Shift & Control Required & Freq.~Dependence \\
\hline
1. & $N \frac{\Delta v}{c} \frac{\omega_a}{c} T^{2} \delta a$ & $\delta a \lesssim {10^{-8} g}/{\sqrt{\text{Hz}}}$ & $\times \left( \frac{\omega / 2 \pi}{10~\text{mHz}} \right)^2$  \\
2. & $N \frac{\Delta v}{c} \omega_a  \delta T$ & $\delta T \lesssim 10^{-12} \text{ s} $ & $\times \left( \frac{\omega / 2 \pi}{10~\text{mHz}} \right)^0$ \\
3. & $N \Delta v \,\delta k\,\Delta\tau  $ & $ c \delta k/2\pi \lesssim 10^2~\text{kHz}/\sqrt{\text{Hz}} $ & $\times\left( \frac{\omega / 2 \pi}{10~\text{mHz}} \right)^0$ \\
4. & $N^2 \frac{\Delta v}{c}\frac{\hbar}{m}\frac{\omega_a}{c} T \delta k$ & $c \delta k /2\pi \lesssim {\text{GHz}}/{\sqrt{\text{Hz}}} $ & $\times\left( \frac{\omega / 2 \pi}{10~\text{mHz}} \right)$ \\
\hline
\end{tabular}
\caption{\label{Tab: phases} A list of dominant noise terms, the control required to achieve a sensitivity of $h \sim \frac{10^{-20}}{\sqrt{\text{Hz}}}$, and the scaling of this requirement with frequency $\omega$.  We assume an example satellite-based configuration with a baseline of 1000 km so the relative velocity between the two atom interferometers is $\Delta v \lesssim 1 \frac{\text{cm}}{\text{s}}$ (see {\it e.g.}~\cite{LISAPPA}).  We take $T \approx 50~\text{s}$, $\Delta\tau \approx 10~\text{ms}$ and $N \approx 300$. All requirements are at a frequency of 10 mHz. These requirements are several orders of magnitude easier to achieve than the state-of-the-art.}
\end{center}
\end{table}

Following the formalism of \cite{Dimopoulos:2008hx} we calculate the differential phase shifts (shown in Table \ref{Tab: phases}) caused by platform acceleration noise $\delta a$,  jitter $\delta T$ in the time between pulses, and laser frequency jitter $\delta k$.
Each of the resulting error terms has its origin only in an initial velocity mismatch $\Delta v$ between the two atomic sources, and is thus suppressed by $\frac{\Delta v}{c} \lesssim 3 \times 10^{-11}$.  In practice, the frequency stability requirements are likely limited by the Rabi frequency associated with the atomic transitions \footnote{For example, frequency errors can lead to pulse area fluctuations which result in an overall loss of contrast.  Keeping contrast loss to 10\% requires $c \delta k / \Omega \sim 0.02$, where $\Omega$ is the Rabi frequency.  For a 1 kHz Rabi frequency, a 20 Hz laser is required.  We note that this is an illustrative limit -- many trades are possible depending on the overall sensor configuration and operating parameters -- {\it e.g.} gravity gradiometer vs. gravity wave detector.}.  Also included in the analysis are corrections related to the finite duration $\Delta\tau$ of the laser pulses \cite{Antoine:2006ApPhB..84..585A}.  The frequency dependence is estimated from the condition $\omega T \sim \pi$ [see Eq.~\eqref{Eq:PhaseShift}], which determines the low-frequency corner of the antenna response \cite{Dimopoulos:2008sv}.  We note that this differential measurement scheme does not remove noise from wavefront aberration \cite{Bender:2011zza, Bender:2011zz}, since after diffraction aberrations are not generally common to both interferometers. However, straightforward noise mitigation schemes suggested in \cite{Dimopoulos:2011zz,Hogan:2010fz} can successfully address these issues.  Finally, ellipse specific methods \cite{Fixler, Foster, Lamporesi} can be used to extract the differential phase shift in the presence of the common-mode laser phase noise.
This is accomplished by operating successive interferometers at a sampling rate higher than the gravitational wave frequency, as described in \cite{Dimopoulos:2007cj, Dimopoulos:2008sv, Hogan:2010fz}.



{\it Atomic Implementation}--  The proposed LMT scheme requires a two-level system with a large (optical) energy difference $\omega_a$ and a long excited state lifetime $\tau$.  To maintain interferometer contrast, the total time $\sim N L/c$ that the atom spends in the excited state during the interferometer sequence cannot exceed $\tau$.  Taking $\tau = N L/c$ as an upper bound, we can write the peak phase sensitivity in Eq. \ref{Eq:PhaseShift} in terms of the quality factor $Q=\omega_a\tau$ of a given atomic transition, resulting in $\Delta\phi_\text{max} = 4\omega_a (N L/c) h = 4 Q h$.  This suggests that the same atoms typically selected for optical clocks because of their high $Q$ transitions are also appropriate for this proposal.  An optical transition with mHz linewidth has $Q> 10^{17}$ which could support a strain sensitivity $h < 10^{-21} /\sqrt{\text{Hz}}$ assuming atom shot-noise limited phase noise $\delta\phi=10^{-4}/\sqrt{\text{Hz}}$.  For gravitational wave detection with $N=300$ and baseline $L=1000~\text{km}$ we have $2N L/c = 2~\text{s}$, requiring at least a sub-Hz linewidth clock transition.

The alkaline earth-like atoms ({\it e.g.}  Sr, Ca, Yb) are promising candidates.  Consider, for example, the clock transition in atomic strontium ($5s^2 \,^1\!S_0 \rightarrow 5s5p\, \,^3\!P_0$). In $^{87}\!\text{Sr}$ this transition is weakly allowed with a linewidth of $1~\text{mHz}$ and a saturation intensity of $0.4 \, \text{pW/cm}^2$ \cite{Takamoto}.  The low saturation intensity enables long-baseline configurations ($>$ 10 km) for suitably cold atomic ensembles \footnote{For high contrast interference, the Doppler widths associated with the velocity spreads of the atomic sources need to be less than the driving Rabi frequency.  In the case where the mean velocities of the two clouds are not matched, two co-propagating frequency components can be tuned to indepenedently address each cloud.}.  In addition to its high $Q$, this transition is also desirable because it exhibits manageable sensitivity to environmental backgrounds.  For example, the blackbody shift has a temperature coefficient of $-2.3~\text{Hz}~(T/300 \text{K})^4$ \cite{Falke}.  At $T = 100~\text{K}$, this implies a temperature stability requirement of $\lesssim 3~\text{mK}/\sqrt{\text{Hz}}$ for a strain sensitivity of $h=10^{-20}/\sqrt{\text{Hz}}$ at 10 mHz.  For magnetic fields, simultaneous or interleaved interrogation of each of the linear Zeeman sensitive transitions, as described in Ref. \cite{Falke}, results in a  residual quadratic Zeeman coefficient of $-0.23~\text{Hz}/\text{G}^2$ \cite{Falke} and also enables measurement of the residual magnetic field.  This coefficient is significantly more favorable than that of the Rb interferometers previously analyzed \cite{Hogan:2010fz}. In principle a second atomic species could be used to independently characterize these shifts in order to provide further suppression.  AC Stark shift related backgrounds appear to be negligible.  Many other backgrounds are similar to those discussed in Refs. \cite{Dimopoulos:2008sv} and \cite{Hogan:2010fz}.

{\it Discussion}--  This configuration enables a high precision measurement of the relative acceleration between two inertial atom clouds.  The high Q atomic transition provides the necessary time reference.  The laser is not used as a clock and thus laser frequency noise does not affect the measurement, unlike all other interferometric gravitational wave detection schemes.  Furthermore, an atom is an excellent inertial proof mass.  A neutral atom's level structure is universal and is significantly less sensitive to environmental perturbations than conventional macroscopic references such as a laser or a drag-free proof mass, whose physical parameters (thermal and electrodynamic properties) can vary significantly.  As we have shown this type of atom interferometer would allow detection of gravitational waves with the same sensitivity as in the proposals described in Refs. \cite{Dimopoulos:2007cj, Dimopoulos:2008sv, Hogan:2010fz} but with significantly reduced requirements on laser and platform stability (as in Table \ref{Tab: phases}), enabling single-baseline gravitational wave detection.

This work was supported in part by NASA GSFC Grant NNX11AM31A.  SR acknowledges support by ERC grant BSMOXFORD no. 228169.  We thank L. Hollberg for useful discussions.

\end{document}